\documentclass[letter]{jpsj2} %% for letters
\begin{document}
\title{Possible Sign-Reversing $s$-Wave Superconductivity in Co-Doped BaFe$_{2}$As$_{2}$ Proved by Thermal Transport Measurements}

\author{Yo \textsc{Machida}, Kosuke \textsc{Tomokuni}, Takayuki \textsc{Isono}, Koichi \textsc{Izawa}, \\Yasuyuki \textsc{Nakajima}$^{1,2}$, and Tsuyoshi \textsc{Tamegai}$^{1,2}$}

\inst{Department of Physics, Tokyo Institute of Technology, Meguro 152-8551, Japan \\
$^{1}$Department of Applied Physics, The University of Tokyo, Bunkyo 113-8656, Japan\\
$^{2}$JST, Transformative Research-Project on Iron Pnictides (TRIP), Bunkyo 113-8656, Japan}

\abst{Thermal transport measurements have been performed on single-crystalline Co-doped BaFe$_{2}$As$_{2}$
down to 0.1 K and under magnetic fields up to 7 T. 
Significant peak anomalies are observed in both thermal conductivity and thermal Hall conductivity
below $T_{\rm c}$ as an indication of the enhancement of the quasiparticle mean-free path.
Moreover, we find a sizable residual $T$-linear term in thermal conductivity,
possibly due to a finite quasiparticle density of states in the superconducting gap induced by impurity pair breaking.
Our findings support a pairing symmetry compatible with the
theoretically predicted sign-reversing $s$-wave state.}

\kword{iron pnictide superconductor, Co-doped BaFe$_{2}$As$_{2}$, thermal transport, sign-reversing $s$-wave state}

\maketitle

The symmetry of the order parameter is essential for identifying the superconducting pairing mechanism.
In conventional superconductors (e.g., Al and Pb), 
the effective electron interaction is mediated by phonons, which gives rise to the isotropic 
$s$-wave pairing symmetry.
On the other hand,
electron pairs glued by magnetic interactions 
form unconventional pairing states: $p$-, $d$-wave states and so on.
So far, such unconventional superconducting states have been found in a
number of materials on the boarder between magnetism and superconductivity~\cite{mathur}. 
These findings suggest that a system close to magnetic instability provides a fertile field
for unconventional superconductivity.
A new family of superconductors containing layers of iron pnictides
bear resemblance to unconventional superconductors such as
high-$T_{\rm c}$ cuprates with a two-dimensional electronic structure and
a magnetic order proximity to the superconducting phase~\cite{kamihara,cruz}.
Therefore, an exotic superconducting pairing state can be naively expected in this system.
In fact, an intriguing pairing state of sign-reversing $s$-wave symmetry has been
theoretically proposed~\cite{mazin,kuroki}.

Here, we report
the first thermal transport evidence of a novel pairing state in Co-doped BaFe$_{2}$As$_{2}$.
In particular,
we find a sizable residual $T$-linear term of the thermal conductivity,
possibly due to the impurity-induced in-gap state.
In addition, significant peak anomalies are observed in both thermal conductivity and thermal Hall conductivity
originating from the prominent enhancement of the quasiparticle (QP) mean-free path below $T_{\rm c}$.
%suggesting the strong AF fluctuation in the normal state.
The field dependence of the delocalized QP density of states is
apparently different from that of nodal gap excitation.
These observations all point to 
the fully gapped sign-reversing $s$-wave state.

Single-crystalline samples with the composition Ba(Fe$_{0.93}$Co$_{0.07}$)$_{2}$As$_{2}$
were grown by the FeAs/CoAs self-flux method~\cite{nakajima}.
The substituted Co atoms donate extra electrons to FeAs layers as itinerant
carriers without creating localized moments~\cite{imai}.
For thermal transport measurements, two different crystals with dimensions of 1 $\times$ 0.4 $\times$ 0.04 mm$^3$ were used.
A one-heater-two-thermometer steady-state method was used to measure thermal conductivity down to
0.1 K and up to 7 T.
Heat current was always aligned within the $ab$ plane of the sample, 
and the magnetic field is applied along the $c$-axis. 
The samples were cooled in a magnetic field to ensure field homogeneity.
We used Cernox and RuO$_2$ thermometers above and below 1 K, respectively.
The thermometers were thermalized on the sample by gold wires held by spot welding,
providing a good thermal contact with a low electrical contact resistance, $R_c \leq$ 50 m$\Omega$, at 300 K.
%We confirm ohmic thermal response with the applied power down to 90 mK
In fact, we confirmed ohmic thermal response with the applied power down to the lowest temperature.
%The gold wires are essential for the stability and the quality of the
%electrical contacts (resistance $R_c \leq$ 50 m$\Omega$ at 300 K), 
%and even more important for reliable thermal contacts between the sample and the thermometers.
The same contacts and gold wires were used
to measure the resistivity of the sample by a
standard four-contact method.

\begin{figure}[t]
\begin{center}
\includegraphics[scale =0.5]{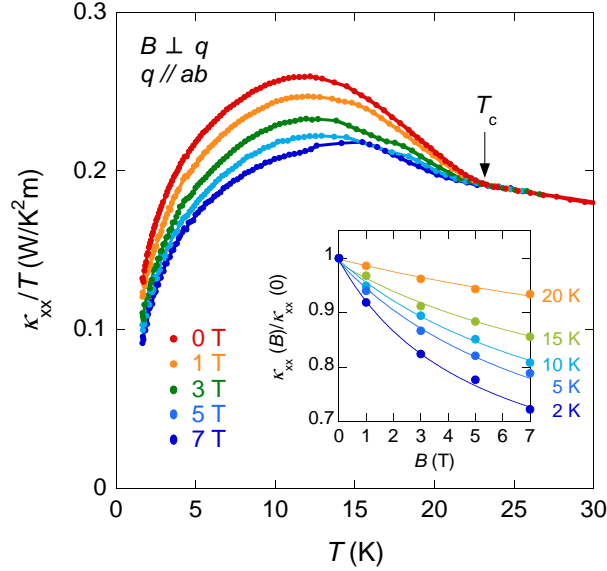}
\end{center}
\vspace{-0.5cm}
\caption{\label{fig:fig.1} (Color online) Temperature dependence 
of thermal conductivity $\kappa_{xx}(T)/T$ under 
several magnetic fields for $B{\perp}q$ and $q{\parallel}ab$.
Inset: Field dependence of normalized thermal conductivity $\kappa_{xx}(B)/\kappa_{xx}(0)$ at 
fixed temperatures. Solid lines are fittings by the vortex scattering model (see text).}
\end{figure}

First, we present the temperature dependence of the thermal conductivity $\kappa_{xx}(T)/T$ 
under several magnetic fields in Fig. \ref{fig:fig.1}.
The arrow indicates $T_{\rm c}$ at zero field determined from the resistivity measurement
shown in the upper inset of Fig.~\ref{fig:fig.4}.
Interestingly, $\kappa_{xx}(T)/T$ increases below $T_{\rm c}$ and reaches a maximum at 12 K ($\sim T_{\rm c}/2$).
In addition, the peak is suppressed by applying a magnetic field.
In general, thermal conductivity is composed of the electronic term $\kappa_{xx}^{\rm e}$
and the phononic term $\kappa_{xx}^{\rm ph}$; $\kappa_{xx} = \kappa_{xx}^{\rm e} + \kappa_{xx}^{\rm ph}$.
Let us estimate $\kappa_{xx}^{\rm e}$ by assuming that the Wiedemann-Franz law is valid at $T_{\rm c}$.
We obtain $\kappa_{xx}^{\rm e}/T = L_0/\rho$ = 1.6 $\times$ 10$^{-2}$ W/K$^2$m $\sim$ 8.4 $\%$ of $\kappa_{xx}/T$,
indicating a predominant phononic contribution above $T_{\rm c}$.
Here, $L_{\rm 0}$ = 2.44 $\times$ 10$^{-8}$ W$\Omega$/K$^2$ is the Lorentz number and
$\rho$ = 150 $\mu\Omega$cm.
Interestingly, a similar increase and its field suppression are observed in hole-doped Ba$_{1-x}$K$_x$Fe$_{2}$As$_{2}$
~\cite{ong} as well as in unconventional superconductors such as YBa$_2$Cu$_3$O$_{7-\delta}$~\cite{YBCO} 
and CeCoIn$_5$~\cite{kasahara}, originating from the
enhancement of the QP mean-free path in the superconducting state.
It should be noted that the enhancement factor ${\kappa_{xx}(T_{\rm c}/2)}/{\kappa_{xx}(T_{\rm c})}$ in Ba$_{1-x}$K$_x$Fe$_{2}$As$_{2}$~\cite{ong}
is two times larger than that in our result.
The origin of this discrepancy will be discussed below.

\begin{figure}[t]
\begin{center}
\includegraphics[scale =0.45]{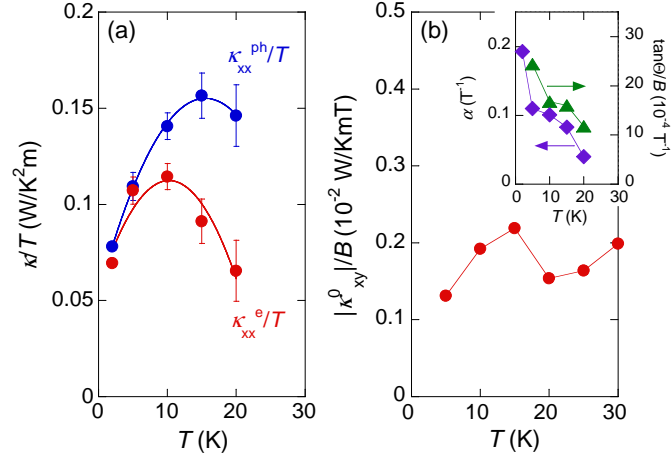}
\end{center}
\vspace{-0.5cm}
\caption{\label{fig:fig.2} (Color online) (a) Temperature dependence 
of electronic term $\kappa_{xx}^{\rm e}(T)/T$ and phononic term $\kappa_{xx}^{\rm ph}(T)/T$
of thermal conductivity.
(b) Temperature dependence of initial slope of thermal Hall conductivity 
${\displaystyle |\kappa^{0}_{xy}|/B {\equiv} \lim_{B{\rightarrow}0}|\kappa_{xy}|/B}$.
Inset: Temperature dependence of $\alpha(T)$ (left axis) and thermal Hall angle ${\rm tan}\Theta/B$ (right axis).}
\end{figure}

To examine the electronic contribution to heat transport,
 we quantitatively estimated $\kappa_{xx}^{\rm e}$ and $\kappa_{xx}^{\rm ph}$
in the superconducting state from the field dependence of the thermal conductivity $\kappa_{xx}(B)$ (inset of Fig.~\ref{fig:fig.1}).
The data can be well described by the vortex-scattering model~\cite{ong,YBCO2}
\begin{equation}
\kappa_{xx}(T,B) = \frac{\kappa_{xx}^{\rm e}(T)}{1+\alpha(T)B} + \kappa_{xx}^{\rm ph}(T),
\label{eq:ong}
\end{equation}
where $\alpha(T)$ is an inverse field scale.
In the analysis,
$\kappa_{xx}^{\rm ph}(T)$ and $\kappa_{xx}^{\rm e}(T)$
are fixed to be field$\nolinebreak -$independent
by assuming that the condensate amplitude is nearly unaffected by magnetic field at $B\leq$ 7 T~\cite{note1}. 
%In this equation, we assume that $\kappa_{xx}^{\rm ph}(T)$ and $\kappa_{xx}^{\rm e}(T)$ are 
%independent of the field, and the dominant
%contribution to the observed field dependence arises from the scattering of QPs by vortices.
%This is because $H^{0}_{\rm c2} \sim$ 30 T~\cite{canfield} is much higher than the applied field ($B \leq$ 7 T), and thus
%the condensate amplitude is nearly unaffected by the field.
Figure~\ref{fig:fig.2}(a) shows the temperature dependences of  $\kappa_{xx}^{\rm e}(T)/T$ and $\kappa_{xx}^{\rm ph}(T)/T$
determined by fitting the data using eq.~(\ref{eq:ong}).
It is noteworthy that $\kappa_{xx}^{\rm e}(T)/T$ exhibits a maximum at around $T_{\rm c}/2$ similar to $\kappa_{xx}(T)/T$.
%while $\kappa_{xx}^{\rm ph}(T)/T$ monotonically decreases below $T_{\rm c}$.
This implies that the peak in $\kappa_{xx}/T$ at zero field is associated
with $\kappa_{xx}^{\rm e}$.
The fittings also provide $\alpha(T)$, which is proportional to the QP mean-free path $l$
through the relation $\alpha(T) = l\sigma_{\rm tr}/\Phi_0$~\cite{YBCO2}, 
where the cross section $\sigma_{\rm tr}$ is roughly equal to the coherence length $\xi$
%where the length scale $\sigma_{\rm tr}$ depends on the particular model 
and $\Phi_0$ is the flux quantum.
In the inset of Fig.~\ref{fig:fig.2}(b) (left axis), 
we show the temperature dependence of $\alpha(T)$, which exhibits a steep increase below $T_{\rm c}$.
This reflects an enhancement of $l$ in the superconducting state.
At 2 K, $l$ is found to be $\sim$ 1100 $\rm{\AA}$
with $\xi$ $\simeq$ 34 $\rm{\AA}$ from $H^{0}_{c2}$ $\sim$ 30 T~\cite{canfield}.

\begin{figure}[t]
\begin{center}
\includegraphics[scale =0.47]{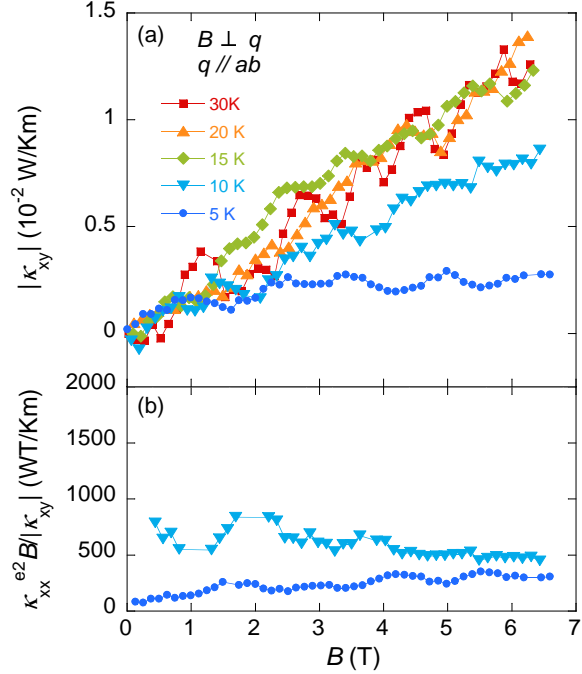}
\end{center}
\vspace{-0.5cm}
\caption{\label{fig:fig.3} (Color online) Field dependence 
of (a) thermal Hall conductivity $|\kappa_{xy}|(B)$
 for $B{\perp}q$ and $q{\parallel}ab$ and 
(b) ${\kappa_{xx}^{\rm e}}^{2}B/|\kappa_{xy}|$ which is proportional to the density of states of delocalized quasiparticles at fixed temperatures.}
\end{figure}

\begin{figure}[t]
\begin{center}
\includegraphics[scale =0.51]{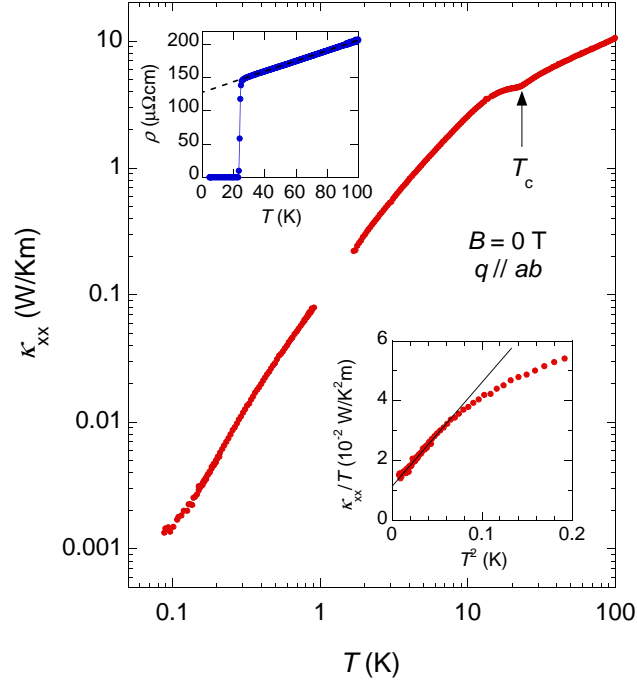}
\end{center}
\vspace{-0.5cm}
\caption{\label{fig:fig.4} (Color online) Temperature dependence 
of thermal conductivity $\kappa_{xx}(T)$ for 
$q{\parallel}ab$ under zero field. 
Upper inset: Temperature dependence of resistivity $\rho(T)$ under zero field.
Lower inset: $\kappa_{xx}/T$ vs $T^2$ plot under zero field.
The solid line represents a fit to the data by $\kappa_{xx}/T=\kappa_{\rm 0}/T + b T^2$.}
\end{figure}

Alternatively, the enhancement of the QP mean-free path
can be confirmed from the thermal Hall conductivity $\kappa_{xy}$,
which is a powerful probe for QPs
since it is purely electronic.
Figure \ref{fig:fig.3}(a) presents the field dependence of $|\kappa_{xy}|(B)$ at fixed temperatures.
The sign of $\kappa_{xy}(B)$ is negative, consistent with that in electron doping
opposite to the hole-doped Ba$_{\rm 1-x}$K$_{\rm x}$Fe$_{2}$As$_{2}$~\cite{ong}.
%$|\kappa_{xy}|(B)$ changes almost linearly against the magnetic field at all measured temperatures.
In Fig. \ref{fig:fig.2}(b), 
we plot the temperature dependence of 
the initial slope ${\displaystyle |\kappa^{0}_{xy}|/B {\equiv} \lim_{B{\rightarrow}0}|\kappa_{xy}|/B}$.
As is clearly seen, $|\kappa^{0}_{xy}|(T)/B$ peaks at around $T_{\rm c}/2$ 
similar to the temperature dependence of $\kappa_{xx}(T)/T$.
This directly indicates the enhancement of the electronic contribution to heat transport below $T_{\rm c}$.
In addition, the QP mean-free path is provided by the thermal Hall angle ${\rm tan}\Theta\equiv|\kappa_{xy}|/\kappa_{xx}^{\rm e}$
because ${\rm tan}\Theta/B\propto l$ in the weak field limit~\cite{kasahara}.
We plot the temperature dependence of ${\rm tan}\Theta/B$ in the inset of Fig. \ref{fig:fig.2}(b) (right axis). 
Clearly, ${\rm tan}\Theta/B$ increases with decreasing temperature,
as we found in $\alpha(T) \propto l$.
Therefore, all data consistently point to the striking enhancement of $l$ below $T_{\rm c}$.
This, in turn, suggests that $l$ in the normal state is suppressed
by the inelastic scattering possibly due to antiferromagnetic (AF) fluctuations, as discussed in the microwave conductivity measurements~\cite{hashimoto}.
On the other hand, in the superconducting state, electrons condense into Cooper pairs 
and their number decreases rapidly below $T_{\rm c}$. 
This in turn gives rise to a reduction in the scattering cross section of QPs, 
and hence $l$ increases below $T_{\rm c}$.
%This, in turn, suggests in the normal state the QPs are significantly scattered, for instance, by the AF fluctuations.
%that we believe to be essential for the $s^\pm$-wave superconducting state.
The presence of AF fluctuations has also been indicated by recent NMR measurements~\cite{imai}.

The thermal Hall conductivity $\kappa_{xy}$ 
also provides the density of states of delocalized QPs, $N_{\rm del}(E)$, using the conjectures
$\kappa_{xx}^{\rm e} \propto N_{\rm del}(E)l$ and $|\kappa_{xy}|/\kappa_{xx}^{\rm e}B \propto l$.
A plot of ${\kappa_{xx}^{\rm e}}^{2}B/|\kappa_{xy}|$ as a function of field reveals the field dependence of $N_{\rm del}(E)$.
Note that the precise estimation of $N_{\rm del}(E)$ from specific heat is rather difficult in iron pnictides
because of the contribution of the nuclear Schottky anomaly~\cite{SH}.
As seen in Fig.~\ref{fig:fig.3}(b), 
${\kappa_{xx}^{\rm e}}^{2}B/|\kappa_{xy}|$ shows a weak field dependence 
within the experimental error of $|\kappa_{xy}|$,
in contrast to the strong field dependence observed in nodal superconductors~\cite{kasahara}.
One may expect such a weak field dependence in fully gapped superconductors because
QPs localized around vortex cores do not contribute to thermal conductivity
at low fields $B \ll H_{\rm c2}$.
%Therefore, the observed field dependence of $N_{\rm del}(E)$ suggests that Ba(Fe$_{0.93}$Co$_{0.07}$)$_{2}$As$_{2}$ is a fully gapped 
%superconductor in accordance with the ARPES measurements~\cite{terashima}.

Next, we discuss the low-temperature thermal conductivity
to gain insight into the superconducting pairing symmetry.
We present the temperature dependence of the thermal conductivity
$\kappa_{xx}(T)$ down to 0.1 K under zero magnetic field in Fig. \ref{fig:fig.4}.
The lower inset of Fig.~\ref{fig:fig.4} shows the low-temperature part of the $\kappa_{xx}/T$ vs $T^2$ plot.
The straight line is a fit to $\kappa_{xx}/T=\kappa_0/T+bT^2$, where $\kappa_0/T$ is the residual $T$-linear term extrapolated to $T\rightarrow$ 0 K.
The best fit was obtained with $\kappa_0/T$ = 1.2 $\times$ 10$^{-2}$ W/K$^2$m and $b$ = 0.33 W/K$^4$m.
What is surprising is that the residual $T$-linear term $\kappa_0/T$ amounts to 
as much as half of the normal$\nolinebreak -$state thermal conductivity $\kappa_{\rm n}/T$,
which is estimated from the Wiedemann-Franz law 
as $\kappa_{\rm n}/T$ = $L_{\rm 0}/\rho_{\rm 0}$ = 1.9 $\times$ 10$^{-2}$ W/K$^2$m.
Here, $\rho_{\rm 0}$ = 130 $\mu\Omega$cm is the residual resistivity at $T$ = 0 K
obtained by assuming that $\rho$ decreases linearly against temperature 
below $T_{\rm c}$ (see dashed line in the upper inset of Fig.~\ref{fig:fig.4}).
%An estimate of the electronic mean
%free path $l$ can be made by the relation
%$\kappa_{\rm n}/T=\gamma v_{\rm F}l/3$
%with the electronic specific heat coefficient $\gamma \sim$ 6 mJ/molK$^2$ and
%the Fermi velocity $v_{\rm F} \sim$ 100 km/s~\cite{dHvA}.
%The determined $l \simeq 70$ ${\rm \AA}$ with $\xi$ $\simeq$ 34 $\rm{\AA}$ from $H^{0}_{c2}$ $\sim$ 30 T
%places Ba(Fe$_{0.932}$Co$_{0.068}$)$_{2}$As$_{2}$ in the moderate clean limit.
On the other hand,
using the mean acoustic phonon velocity $\langle v_s \rangle$ = 2400 m/s 
and the phonon specific heat coefficient $\beta$ = 8.98 J/K$^4$m$^3$
obtained from the parent compound of BaFe$_2$As$_2$~\cite{sefat}, 
the slope of $b$ = $\frac{1}{3}{\beta}{\langle}v_s{\rangle}l_{\rm ph}$ yields $l_{\rm ph}$ = 46 $\mu$m,
which is the same order of magnitude as the smallest crystal dimension, namely, 40 $\mu$m thickness of the sample. 
This implies that the low-temperature thermal conductivity is dominated by phonons.
%Therefore, the $T^3$
%term in $\kappa_{xx}$ originates from phonons.

In the superconducting state, the presence of the residual $T$-linear term $\kappa_{\rm 0}/T$ can be
attributed to the impurity bound states
formed by the
interference of particle- and hole-like excitations, which
undergo Andreev scattering, evoking sign changes of 
the order parameter as a result of unconventional pairing and impurity scattering.
%Therefore,
%the density of Andreev bound states is finite for finite
%concentration of  impurities.
%As a consequence, this gives rise to a residual $T$-linear term of the thermal conductivity.
Moreover, in the nodal superconductors, $\kappa_{\rm 0}/T$ takes
a universal value independent of impurity concentration because
the quasiparticles are both generated and scattered by the impurities~\cite{graf}.
In fact, the impurity concentration independent $\kappa_{\rm 0}/T$ 
has been observed in high-$T_{\rm c}$ superconductors~\cite{taillefer},
%where the $d$-wave symmetry with lines of nodes is firmly established,
although its universality is still under debate~\cite{ando}.
Theoretically, the universal thermal conductivity $\kappa_{\rm 00}/T$ for the
nodal superconductor is explicitly expressed as
\begin{equation}
\frac{\kappa_{\rm 00}}{T} = \frac{\pi^2}{3}N_fv_f^2\times\frac{a\hbar}{2\mu\Delta_0},
\end{equation}
%\begin{equation}
%\frac{\kappa_{\rm 00}}{T} = \biggl(\frac{\pi}{4}\frac{\hbar\Gamma}{\Delta_0}\frac{1}{\mu}\biggr)\frac{\kappa_{\mathrm n}}{T},
%\end{equation}
where 
$N_f$ is the normal density of states,
$v_f$ is the Fermi velocity,
$\Delta_0$ is the maximum
amplitude of the gap, $\mu$ is the slope of the gap at the
node on the circular Fermi surface, and
$a$ is $\frac{4}{\pi}$ for the 2D order parameter with lines of nodes~\cite{graf}.
Using the Wiedemann-Franz law,
eq. (2) can be written as $\frac{\kappa_{\rm 00}}{T} = \bigl(\frac{\pi}{4}\frac{2l\pi}{\xi}\frac{1}{\mu}\bigr)\frac{\kappa_{\mathrm n}}{T}$~\cite{izawa}.
By assuming a $d$-wave gap structure,
we obtain $\kappa_{00}/T$ = 0.4 $\times$ 10$^{-3}$ W/K$^2$m for Ba(Fe$_{0.93}$Co$_{0.07}$)$_{2}$As$_{2}$
using $\mu$ = 2 with $\xi$ = 34 $\rm{\AA}$ and $l$ = 1100 $\rm{\AA}$.
%using $\hbar\Gamma/\Delta_0 \simeq \pi\xi/2l \sim$ 0.06 and $\mu$ = 2 with $\xi$ = 34 $\rm{\AA}$ and $l$ = 1100 $\rm{\AA}$.
The determined value is one order of magnitude smaller than the experimental value of $\kappa_{\rm 0}/T$.
Although we cannot exclude the possibility of the nodal superconducting state on the basis of only $\kappa_{00}/T$, 
given the leading phononic contribution in low-temperature $\kappa_{xx}$, as evidenced by the $T^3$ term
and the weak field dependence of $N_{\rm del}$,
nodal superconductivity is highly unlikely in Ba(Fe$_{0.93}$Co$_{0.07}$)$_{2}$As$_{2}$.
Here, a question is raised as to how one can take a finite $T$-linear term in $\kappa_{xx}/T$ without the nodal pairing state.
%suggest the fully gapped state in Ba(Fe$_{0.93}$Co$_{0.07}$)$_{2}$As$_{2}$.

In addition to the nodal superconducting state,
one can also expect a residual $T$-linear term in a novel $s$-wave pairing state.
At first glance, the residual $T$-linear term seems to be incompatible with the $s$-wave state on the basis of Anderson's theorem~\cite{anderson}.
On the contrary, it is possible to realize the finite value via impurity interband scattering between different Fermi surfaces with opposite signs of order
parameters with full gaps.
%This can be realized by
%reversing the sign of order parameter between the opposite sign of the $s$-wave gaps on the different Fermi surfaces via
%the impurity interband scattering.
In fact, it is theoretically proposed that
the AF fluctuations with ${\bf Q} \approx (\pi,0)$ resulting from the interband
nesting between the hole and electron pockets 
realize a fully gapped sign-reversing $s$-wave state ($s_{\pm}$-wave state) in the iron pnictide superconductors~\cite{mazin,kuroki}.
In this state,  the impurity-induced density of states (DOS) in the gaps is predicted to emerge
when interband scattering is enhanced by the 
introduction of impurities~\cite{kontani}.
However, it should be emphasized that such an in-gap state can be expected only if
a strong impurity pair breaking involving a large reduction in $T_{\rm c}$, which amounts to 10 K, occurs~\cite{kontani}.
% a large reduction in $T_{\rm c}$, which amounts to 10 K,
%occurs due the strong impurity pair-breaking~\cite{kontani}.

Here, we discuss whether such a large reduction in $T_{\rm c}$ occurs in Co-doped BaFe$_{2}$As$_{2}$
to examine the possibility of the $s_{\pm}$-wave state
in this system.
%The AF fluctuations, which is essential for the $s_{\pm}$-wave state,
%is indicated by the enhancement of the QP mean-free path in the superconducting state.
Given that $T_{\rm c}$ = 25 K of Co-doped BaFe$_{2}$As$_{2}$ is about 10 K lower than
that of the K-doped one, $T_{\rm c}$ = 37 K~\cite{ong},
the difference in $T_{\rm c}$ can be explained by the difference in the strength of the pair breaking caused by the dopant.
%$T_{\rm c}$ is reduced possibly due to the impurity pair-breaking.
This is because the Co atoms substituted into the conducting layer can act as stronger pair breakers than the K atoms doped into the
block layer.
In fact, 
%The presence of strong scattering by the Co atoms is also suggested by
the enhancement of the thermal conductivity ${\kappa_{xx}(T_{\rm c}/2)}/{\kappa_{xx}(T_{\rm c})}$  for Ba(Fe$_{0.93}$Co$_{0.07}$)$_{2}$As$_{2}$
is reduced to half of that for K-doped BaFe$_{2}$As$_{2}$~\cite{ong}.
In addition, the resistivity at $T_{\rm c}$  for Ba(Fe$_{0.93}$Co$_{0.07}$)$_{2}$As$_{2}$ ($\rho$ = 150 $\mu\Omega$cm) is three times larger
than that for the K-doped one ($\rho$ = 50 $\mu\Omega$cm)~\cite{ong}.
These results indicate that the scattering rate of Ba(Fe$_{0.93}$Co$_{0.07}$)$_{2}$As$_{2}$ is practically 
enhanced by Co doping.
%Furthermore, suppose $T_{\rm c}$ is reduced by 10 K due to 
%Furthermore, if the difference of $T_{\rm c}$ can only account for the impurity pair-breaking by the Co atoms,
In this case,
one may expect the absence of an in-gap state or a smaller in-gap state in K-doped BaFe$_{2}$As$_{2}$ with weak scattering by the K atoms.
%Therefore, $T_{\rm c}$ is practically reduced, and thus the impurity induced in-gap DOS can be expected.
%As the result of impurity pair-breaking by the Co atoms, the finite DOS in the gaps is
%induced and leads to the sizable residual $T$-linear term in the Co-doped BaFe$_{2}$As$_{2}$. 
%the impurity bound state is formed and leads to
%a finite in-gap DOS. 
%This eventually provides the residual $T$-linear term of $\kappa_{xx}/T$.
Recently, a negligible residual $T$-linear term in $\kappa_{xx}/T$ was found in Ba$_{1-x}$K$_x$Fe$_{2}$As$_{2}$~\cite{luo}.
%implies the role of the Co atoms doped in the conducting layer for the finite in-gap DOS.
Moreover, the spin-lattice relaxation rate $1/T_1$ of the Co-doped BaFe$_{2}$As$_{2}$ 
appears to level off at low temperatures~\cite{imai}, while that of the K-doped one varies close to $T^3$
all the way down to the lowest measured temperature~\cite{fukazawa}.
These results further explain the role of Co atoms as strong pair breakers and support our scenario.
%These results also suggest the presence of residual DOS in the the Co-doped BaFe$_{2}$As$_{2}$ induced by the Co doping.
Thus, our findings all point to the fully gapped $s_{\pm}$-wave state 
in Ba(Fe$_{0.93}$Co$_{0.07}$)$_{2}$As$_{2}$,
being consistent with the theoretical prediction as well as the resonance peak observed by the
inelastic neutron scattering measurement~\cite{INS}.

In summary, we uncovered an intriguing pairing state of the fully gapped sign-reversing $s$-wave state 
in the Co-doped BaFe$_{2}$As$_{2}$ 
by thermal transport measurements.
Further experiments on samples with different Co doping
levels are in progress
to clarify the effect of inhomogeneity in the sample and
the crucial role of non-magnetic impurities to the residual in-gap state.

We thank K. Hirata for providing experimental equipment
and K. Behnia, H. Kontani, and Y. Matsuda for useful discussions.
This work is supported in part by Grants in Aids (No. 20684016, 20840015) 
for Scientific Research  from the Japanese Society for the Promotion of Science,
by a Grant-in-Aid for Scientific Research on Innovative Areas "Heavy Electrons" (No. 20102006) 
of The Ministry of Education, Culture, Sports, Science, and Technology
and by the Global Center of Excellence Program from MEXT through the Nanoscience and
Quantum Physics Project of the Tokyo Institute of Technology.

\end{document}